# Moving Away from the Joneses to Move Ahead: Migration, Information Gap and Signalling

*Shihas Abdul-Razak, Upasak Das, and Rupayan Pal*[1]

March 27, 2023

**Abstract**

*Signalling social status through the consumption of visible goods has often been perceived as a way in which individuals seek to emulate or move up compared to others within the community. Using representative migration survey data from the Indian state of Kerala, this paper assesses the impact of transnational migration on consumption of visible goods. We utilize the plausibly exogenous variation in migration networks in the neighbourhood and religious communities to account for the potential endogeneity. The findings indicate a significantly positive and robust effect of migration on conspicuous consumption, even after controlling for household income. In terms of the mechanisms, while we are unable to rule out the associated taste-based changes in preferences and the peer group effects driving up the spending on status goods, we observe only limited effects of these channels. A potential channel that we propose is information gap among permanent residents about the income levels of an out-migrant, which is leveraged by them to signal higher status in society. We explore this channel through a theoretical model where we connect migration, information gap and status good consumption. We derive a set of conditions to test whether migrants exhibit snobbish or conformist behaviour. Empirical observations indicate predominance of a snobbish behaviour.*

## 1 Introduction

Social status within a community remains one of the key elements influencing human behaviour. Veblen in his pioneering work, "*The Theory of the Leisure Class*", argues "the members of each stratum accept as their ideal of decency the scheme of life in vogue in the next higher stratum, and bend their energies to live up to that ideal" (Veblen, 1899). Duesenberry (1952) builds on this premise by advancing the discourse on social signalling and status. One common way of signalling social status is the consumption of visible goods or conspicuous consumption, which is often perceived as how individuals define themselves in society. As a method of household reproduction in society, households seek to emulate or surpass the consumption patterns of their social group or other groups, especially when unequal social groups co-exist (Carrier & McC, 1997).

---

[1] Shihas Abdul-Razak: Yale University, shihas.abdulrazak@yale.edu; Upasak Das: Global Development Institute, University of Manchester, upasak.das@manchester.ac.uk; Rupayan Pal: Indira Gandhi Institute of Development Research (IGIDR), Mumbai, rupayan@igidr.ac.in. An earlier version of this paper was circulated as "Migration, Information Gap, and Visible Consumption: Evidence from India." We thank Soham Sahoo, Linh To, and conference and seminar participants at Delhi Winter School 2020, IIM Bangalore and CDS Thiruvananthapuram for their helpful comments and feedback.

Existing studies have empirically demonstrated the existence and heterogeneity of conspicuous consumption in different contexts. For example, literature has shown how a household's socioeconomic position within their own reference group and the self-esteem of the individuals get associated with different levels of visible consumption, which is found to vary across different groups as well (Bertrand & Morse, 2016; Bursztyn et al., 2017; Charles et al., 2009; Kaus, 2013; Khamis et al., 2012). Signalling through consumption is also found to have a bearing on portfolio choices in the financial markets and household indebtedness (Bricker et al., 2020; Gali, 1994; Roussanov, 2010). In addition, studies have documented how inequality in the reference and socio-economic groups affect conspicuous consumption (Frank et al., 2014; Roychowdhury, 2016). This paper looks at a different aspect: whether transnational migration of households can influence conspicuous consumption and social signalling. In the process of assessing the impact, we develop a theoretical model to connect migration, information gap and status signalling. The study uses representative household migration data collected in 2018 from the Indian state of Kerala. The case of Kerala is important because of the historically high prevalence of out-migration primarily to the Middle-Eastern countries for low and semi-skilled employment.

Using a basic regression framework, we first examine the effects of transnational migration on conspicuous household consumption. To examine the sensitivity of the estimates under the presence of the unobserved heterogeneity because of the self-selection of households into migration, we estimate the bias adjusted treatment effect using the coefficient stability method developed by Altonji et al. (2005) and operationalized by Oster (2019). Further, for identification, we use exogenous variation in the proportion of households of the same religion within an aggregated geographical area having current migrants as an instrument variable (IV). In addition, we also test whether the stability of the estimates with substantial relaxation of the assumption of strict exogeneity using methods developed by Conley et al. (2012) and van Kippersluis & Rietveld (2018). We also use Inverse Probability Weighting Regression Adjustment (IPWRA) that gives higher weightage to households with lower chances of being treated/untreated to confirm that the results are robust.

Our findings indicate that migrant households are more likely to spend on conspicuous consumption and possess luxury house in comparison to non-migrant households. This is after controlling for a range of observable confounders. Using the methods elucidated above, our estimations are robust to selection due to unobservables indicating a potential causal relationship between transnational migration and spending on visible consumption.

What is the potential relationship between transnational migration and social signalling? In other words, why would a migrant have different levels of the propensity of visible consumption compared to a non-migrant? We explore two channels potentially responsible for the systematically higher spending observed among migrants. Firstly, the difference in the consumption basket of migrants and non-migrants could be driven by taste-based changes in the migrants' consumption demand post-migration. We present evidence to show that transnational migration influences total consumption expenditure, though it includes both visible and other expenses. Our findings indicate no suggestive evidence of the possibility of taste-based changes driving the observed higher visible consumption among migrants, though we do not rule out this possibility. Secondly, we argue visible consumption among migrants might be driven by peer effects wherein reference group income level, or inequality may drive the visible good consumption. This emanates from the literature that documents reference or peer group effects on conspicuous consumption where reference group income level and visible inequality are systematically associated with status good possession and spending (Charles et al., 2009; Roychowdhury, 2016). To test this possibility, we add indicators that capture reference group income distribution as control variables in the main regressions, however, we do not find any substantial changes in the migration effect size. This indicates the reference group effect also do not play a role in explaining the impact of transnational migration in our context.

The third argument we propose is increased incentives for status signalling through differences in information gaps. Existing studies indicate prevalence of a significant gap in the information among the home community about a migrant's job profile and actual wealth status while migrating outside (Baseler, 2021; Joseph et al., 2018). This higher level of information gap may incentivize the migrants to seek status by increasing visible consumption, which can be leveraged to signal income levels. To argue this, we use qualitative insights from a set of interviews conducted with 20 migrants and return migrants from two districts of Kerala. We find strong evidence of substantial information gap on the employment details and earnings of the migrants among their household members as documented in the literature (Seshan & Zubrickas, 2017). Using the findings from the literature and the qualitative interviews, we use the framework of Corneo & Jeanne (1997) to develop a theoretical model to connect migration with the information gap and signalling through visible good consumption. We consider a society with two types of households, migrant and non-migrant, wherein higher income brings a higher level of social status. We demonstrate that in the equilibrium under asymmetric information regarding migrant households' income, the propensity to spend on status goods is higher for migrant households compared to their non-migrant counterparts. Unlike in the case of non-migrant households, the signal value of status good consumption to a migrant household is positive and greater than the opportunity cost of status good consumption, unless the migrant household's actual income is sufficiently low. Next, the prevailing social norm may be such that the gain in (loss of) social status by climbing up (sliding down) one position in the income hierarchy is more for a higher (lower) ranked household compared to that for a lower (higher) ranked household. That is, when more households can identify with the rich, the signal value of status good consumption may be lower (snob effect) or higher (conformity effect), depending on the prevailing norm in the society. We derive a set of conditions to test whether status-good consumption exhibits snobbish or conformist behaviour in the present context. Empirical evidence from the study data shows a predominance of snobbish behaviour on average exhibited by the migrants based on status good consumption.

The paper primarily contributes to three strands of literature. Firstly, it adds to the literature on the economics of conspicuous consumption. In this context, we contribute by relating migration with social status signalling through visible consumption. Secondly, it contributes to the literature on transnational migration and remittances and their effects on different dimensions of consumption. Existing literature has documented significant gains on education, health and income (Antman, 2012, 2013; Anyanwu & Erhijakpor, 2010; Hildebrandt et al., 2005; Ratha et al., 2011; Stillman et al., 2009). In this paper, we present a strong and robust impact on visible consumption. Three, it contributes to the literature on information asymmetry and argues how individuals can leverage the differences in the level of imperfection in information about actual income/wealth to ensure higher social status in the community. Here, we develop a formal analytical framework to explain the implications of this information gap being potentially utilized by transnational migrants to signal social status in the home country. Evidence from our study data indicates that the migrants exhibit snobbish behaviour. The findings indicate that information gap heterogeneity plays a crucial role in determining the level of conspicuous consumption, and this has broader implications that go beyond the migrants. With an expansion of the economy and changing economic roles, mobility, and employment opportunities, the returns from conspicuous consumption might also increase disproportionately. This would have larger welfare consequences as more resources would be diverted to the consumption of visible items (Frank, 2005). Higher incentives for conspicuous consumption due to larger information gap could explain the increasing consumption and consumer debt in the economy (Leicht, 2012; Weller, 2007), which in turn potentially have macroeconomic consequences (Jordà et al., 2016; Mian et al., 2017).

The rest of the paper is organized as follows. The next section provides the background on migration from Kerala and its distinctiveness. Section 3 explains the data and empirical strategy. The results of the empirical analyses are presented in Section 4. Section 5 explores the relationship between conspicuous consumption and the information gap formally. The last section provides some concluding remarks.

## 2 Kerala and Migration

Kerala, a state in the Southern part of India, is renowned for its high levels of human development and the peculiar developmental path it has gone through (CDS-UN, 1975; Isaac & Tharakan, 1995). Though the history of migration in the state dates back to time immemorial, there has been a systemic increase in migration from Kerala to foreign countries in search of jobs from the 1970s. Most of this outflow has been for unskilled and semi-skilled jobs in Middle-Eastern countries, primarily Bahrain, Kuwait, Oman, Qatar, Saudi Arabia, and the United Arab Emirates. This was in response to the increased labour demand in the region that started from the 1970s owing to the oil boom. This migration was temporary, and the migrant workers always had the idea of returning to the homeland. In 2011, approximately 17% of Indian migrants living abroad were from Kerala.[2] Most of these migrants lived alone without family in labour camps and temporary residences (Seshan, 2012; Zachariah et al., 2004). The fact that most of the migrants left their families back in Kerala, along with the temporary nature of migration, led to the inflow of a significant chunk of migrants' earnings to the state's households in the form of remittances (Raman, 2015; Zachariah et al., 2002). The systemic migration and the subsequent foreign remittance flows, which account for almost 20% of the State Domestic Product (Rajan & Zachariah, 2020), have changed the socioeconomic contour of the state. Remittances have been considered the major factor contributing to the increased economic growth of the state from the 80s (Kannan, 2005; Kannan & Hari, 2020). Kerala has also been considered as the vanguard of consumption in India, which is majorly driven by foreign remittances. Scholars have also noted the consequences of the temporary nature of the migration on the financial decisions of the households in the state (Gulati, 1993; Osella & Osella, 1999). In addition, migration and remittances inflow have impacted wide range of other socio-economic aspects in the state, such as labour force participation (Khan & Valatheeswaran, 2016) and health (Ali et al., 2017) of left-behind members, private schooling (Valatheeswaran & Khan, 2018), and asset accumulation (Seshan, 2020).

[2] Calculated by authors using estimates from Zachariah & Rajan (2012) and UN-DESA (2020).

The peculiar features of the Kerala migration and the cross-country migration of low or semi-skilled labours from developing regions in general make it an appropriate setting to explore the link between migration and visible consumption. Firstly, unlike other distress-induced unskilled and semi-skilled migration, international migration to the Middle-East from South Asia did not accompany reputational damage. In Kerala, these migrants were regarded as fortune-bringers and considered with equal, if not higher, status in society (Dominic, 2017; Nidheesh, 2020). Migrants are often seen as the driving force behind the consumption trends in the state, and the consumption among migrants appealed to many young people and made them migrate (Pattath, 2020). The second feature is the temporary nature of the migration. The Middle-Eastern countries never provided the migrants with an option of settling down there permanently. As we indicated above for majority of these migrants, their family members always remained back in Kerala.

## 3 Data and Empirical Strategy

### 3.1 Data

Data used for this study comes from Kerala Migration Survey (KMS) conducted in 2018. KMS is a periodic household survey conducted by the Centre for Development Studies, Thiruvananthapuram, to monitor migration from Kerala. This survey data is a unique resource for migration studies, especially given the minimal availability of reliable data on migration globally. The first round of KMS was conducted in 1998, and since then there have been seven more rounds, with the latest data available round in 2018. The 2018 round dataset covers 15000 households across the state. The sample households are selected using stratified multistage random sampling, where district rural and urban areas are the strata. The number of sample households in each stratum is selected in proportion to the census data. One ward is chosen randomly from a selected number of localities (village/municipality). From each ward selected, 30 households are chosen randomly for the survey. The number of localities in each stratum is decided by the number of households needed in the stratum. In 2018, there are 445 localities divided proportionally in the entire sample.[3]

### 3.2 Variables

To capture visible consumption, we use possession of a luxurious house as our primary dependent variable. Marriages and houses are considered the primary means through which social status is often displayed and gauged (Osella & Osella, 1999). For our study, luxury houses are defined as those that have at least two bedrooms with attached bathrooms with tile flooring (see Table 1). In addition, we use two other indicators of visible consumption from the data on possession of goods with high visibility as the dependent variables. First, we follow the categorization of goods done by Khamis et al. (2012).[4] This categorization gives a list of goods/assets[5] whose possession can be characterized as highly visible. Notably, the KMS has collected data on possession of ten such goods. We find the least-possessed five goods for each locality out of these ten assets. If a household possesses at least one of the five least-possessed goods in its locality, it is categorized as one spending on conspicuous consumption. Accordingly, it is given a value of 1 but 0 otherwise. To construct the second indicator, we follow three steps. First, for each of the ten assets categorized in Khamis et al. (2012), we assign a value of 1 to a household if it possesses the respective asset and 0 if it does not. Therefore, for each household, we create a set of ten binary indicators that capture possession of each of the ten assets. In the second step, we standardize these ten binary variables and add them for a household to get a Visible Consumption Index (VCI). Finally, we standardize the sum value to obtain the standardized VCI. These types of indices have been used previously in the literature to derive empowerment score for women (Heath & Tan, 2020).

[3] For the cities (municipal corporations), the sampled households are more than 30 to represent the larger geographical area and population. For more details on the sampling method, see Zachariah & Rajan (1997).

We capture migration through a dummy variable that identifies if a household has at least one international migrant. This constitutes our primary variable of interest. We control for a range of household demographics and characteristics as control variables. It is possible that the marginal propensity of visible consumption increases with income and hence economic condition of the household is incorporated. Therefore, we use the inverse hyperbolic sine value of household income to account for the income effects on visible consumption. Further, different demographic groups may have different propensities to consume visible goods, and demographics also can influence the incentive for status-seeking (Charles et al., 2009; Khamis et al., 2012). Accordingly, we control for religion, area of living, highest education within the household, and type of poverty card owned.[6] To control for the differences in household sizes, we use the number of members in the household along with the number of earning members. We also control for the regional effects (Mejia & Restrepo, 2016) by including district dummies.

Table A1 presents the summary statistics of all these variables separately for the non-migrant and the migrant households. We also add a third column here to show the difference between these variables between migrant and non-migrant households and test whether these differences are statistically indistinguishable from zero. The findings reveal that migrant households have a significantly higher consumption of visible goods than non-migrant households. For example, 61 percent of the migrants own a luxury house compared to 39 percent among the non-migrants and the difference is statistically significant at 1% level.

---

[4] Khamis et al. (2012) conducts a survey, following Charles et al. (2009) and Heffetz (2011), in India to find out the goods that possess high visibility compared to other goods.
[5] See Table 1 for the list of these ten assets.
[6] The poverty card categorizes households by their poverty status and is used by the government for its social security programs. The categories are shown in Table A1.

**Table 1**: Definition of Visible Consumption Indices

Type of house which the household is now occupying

Coding of responses:
1 - 3 or more bedrooms with attached bathrooms, concrete/tile roof, tiled floor
2 - 2 bed rooms with attached bathrooms, concrete/tile roof, Mosaic floor
3 - 1 bed room, brick and cement walls, concrete or tile roof
4 - Brick walls, cement floor, tin or asbestos roof
5 - Mud walls, Mud floor & Thatched roof

**Luxury House** – Response 1 or 2

---

Does the household possess the following?

Coding of responses: 0 - No, 1 – Yes

(1) Motorcar
(2) Motorcycle
(3) Mobilephone
(4) Television
(5) Refrigerator
(6) Washing Machine
(7) Oven
(8) Computer
(9) Air Conditioner
(10) Inverter

**Conspicuous Consumption Household** – Household possesses at least one of the five least possessed goods in the region (Taluk)

**Visible Consumption Index** – After taking standardized value for each of these ten goods, a sum total of these ten standardized scores is taken for each household, which is again standardized

---

### 3.2 Empirical Strategy

Guided by several studies in the Indian context that explain consumption pattern, our baseline econometric model is given by:

$$Y_{id} = \alpha + \beta D_{id} + \phi X_i + \delta_d + \varepsilon_i \qquad (1)$$

where $Y_{id}$ gives the measure/indices of visible consumption as discussed for household $i$ from district, $d$. In particular, $Y_{id}$ takes the value of 1 if the household possess a luxury household and 0 otherwise. In other models, we first take $Y_{id}$ to be the standardized VCI. We also consider $Y_{id}$ as a binary variable to categorize those with five least possessed visible consumption goods in the locality. $D_{id}$ takes the value of 1 if the household has at least one member who is an international migrant and zero otherwise. In another specification, it takes the inverse hyperbolic sine transformation value of the remittances. $\beta$, which gives the estimated treatment effect, is our coefficient of interest. $X_i$ is the vector of observed covariates including household income. The district-fixed effects are given by $\delta_d$. $\varepsilon_i$ is the error term.

The above regression can yield unbiased estimates if the migration decision of the household and remittances are exogenous to visible consumption or it is conditionally exogenous solely through the observable characteristics. However, if migration is determined by unobserved heterogeneities that also confound visible consumption, our estimates would be biased. To address this, we use the coefficient stability test by Altonji et al. (2005) and Oster (2019), which provides the conditions on the unobservable characteristics under which the results from models with observed controls hold. It assumes that the relationship between the treatment and the observables are informative about the relationship between the treatment and unobservables. The method provides a strategy to generate bounds of the treatment coefficient accounting for the effect of unobservable characteristics.

In addition, we employ IV regression, where the decision to migrate is instrumented by variation in the prevalence of international migrants within the religious group in the *taluk*[7] of the corresponding household. Notably, to gauge the effects of migration, literature has used the prevalence of migration within the community as IVs as this would be dependent on the pre-existing migration network and hence highly correlated with the household migration decision (Amuedo-Dorantes & Pozo, 2010; Bouoiyour & Miftah, 2016; Datt et al., 2020). In our case, it must be noted that locality-specific migration networks often are the key reasons that explain variation in migration across households and regions. In 2010, for example, surveys found 78% of the migrant households knew at least one relative, friend, or neighbor who had emigrated outside India to work, which is only 43% for the non-migrants.[8] In our case, the IV denotes the prevalence of previous migrants within one's own religious group and the *taluk*. Having higher former migrants in the community is likely to predict an increased migration network among the residents and hence a higher likelihood of current international migration.

[7] *Taluk* (or *Tehsil*) is a local administrative subdivision of a district consisting of a number of villages and towns.
[8] The figures are calculated from Kerala Migration Survey (KMS) 2010 round data.

The reliability of the IV results is based on the assumption that the IV is uncorrelated with the unobserved characteristics driving household visible consumption through channels other than that from international migrants within the group. However, this assumption can be questioned on several grounds. For instance, if migration has an impact on visible consumption, then a higher migration network could impact visible consumption of non-migrants through peer effect. Therefore, we relax this assumption and allow the instruments to be plausibly exogenous. In other words, we use the IV regression method developed by van Kippersluis & Rietveld (2018) based on Conley et al. (2012) that preserves the causal estimation but with weaker instruments. With weaker assumptions, Conley et al. (2012) show that the bounds rather than the point estimations can be obtained from the endogenous parameter of interest. To estimate point estimates, van Kippersluis & Rietveld (2018) build on this by exploiting zero-first-stage sub-populations present in the sample. In other words, by using van Kippersluis & Rietveld (2018) and Conley et al. (2012), we can get the econometric framework to understand the implication of migration on visible consumption when the migration network is imperfect instrument, and may only be plausibly exogenous.

According to this, we consider the equation below:

$$Y_{id} = \beta D_{id} + \gamma Z_{id} + \delta_d + \phi X_{id} + \varepsilon_i \quad (2)$$

Where $Y_{id}$ is the measure of visible consumption for household, $i$. $D_{id}$ refers to whether the household has an international migrant and is endogenous in nature, $Z_{id}$ is the IV as indicated and $X$ is the vector of the control variables. The IV regression estimates assume that $\gamma$ =0. However, if the parameter $\gamma$ is close to zero, but not necessarily zero, it can be said that the instrument is plausibly exogenous and we can still consistently estimate the effect of transnational migration on the visible consumption for certain values of $\gamma$. Existing literature has used this methodology in various contexts (Azar et al., 2022; Biswas & Das, 2022; Das, 2021; McArthur & McCord, 2017).

## 4 Results and Discussion

### 4.1 Baseline results

We first discuss the estimates from the regression to estimate the effect of having migrants on the indices of visible consumption. Here we present the results to show the impact of having a transnational migrant. As discussed, we use three outcome variables: possession of luxury houses, possession of at least one among five least possessed assets within the locality and standardized VCI. Table 2 presents the regression estimations. Because the first two outcome variables are binary in nature, the marginal effects from the probit regressions are also shown.

**Table 2**: Effect of transnational migration on visible consumption

| | Luxury House | | 5 least possessed | | VCI | |
|---|---|---|---|---|---|---|
| | (1) | (2) | (3) | (4) | (5) | (6) |
| Migrant Household | 0.546*** (0.038) | 0.433*** (0.040) | 0.505*** (0.034) | 0.358*** (0.039) | 0.447*** (0.029) | 0.228*** (0.023) |
| Controls | No | Yes | No | Yes | No | Yes |
| Marginal Effect | 0.210*** (0.014) | 0.134*** (0.012) | 0.184*** (0.012) | 0.099*** (0.011) | | |
| Observations | 14325 | 14325 | 14325 | 14325 | 14325 | 14325 |
| $\hat{\delta}\ for\ \beta = 0$ | | 3.061 | | 2.383 | | 2.040 |

*Note. Estimations from probit regression are presented in column (1) to (4) and for column (5) and column (6), those from OLS regression are presented. The numbers in the parenthesis are the standard error clustered at the locality level. The dependent variables are as follows (i) column (1) and (2): possession of luxury house, (ii) column (3) and column (4): possession of at least one out of five least possessed goods in the locality and (iii) column (5) and column (6): visible consumption index (VCI). The main variable of interest is whether the household has at least one international migrant. The control variables are those listed in Table A1 and district fixed effect.* $*** \ p < 0.01$; $** \ p < 0.05$; $* \ p < 0.1$. *In the last row, the value of* $\hat{\delta}$, *from the coefficient stability (Oster) test, for which the coefficient estimate turns zero is given. Detailed results with all coefficients are presented in Table A2 and detailed results from coefficient stability test is given in Table A3.*

The findings indicate a positive effect of having international migrants in the households on visible consumption, which is found to be statistically significant below 5% level. In terms of the effect size, we find that having a transnational migrant in the household increases the likelihood of possessing a luxury house by 13 percentage points. Similarly, the chance of possessing at least one of the five least possessed goods increases by 10 percentage points. We also find an increase in the VCI by 0.23 standard deviations on average. Note that we have controlled for a robust set of other economic factors, including household income, the number of earning members, highest education in the household and ration card type, among others. Hence, we argue that international migration potentially creates additional incentives among the migrant households to spend more on consumption items that are visible in nature.

The full regression estimates with the control covariates are reported in Table A2. Expectedly, household income has a secular positive effect on visible consumption. Christians and Muslims are more likely to possess visible goods than Hindus. Higher education is also associated with higher possession of such goods. Rural households are less likely to possess such goods compared to their urban counterparts.

*Coefficient Stability Test*

The credibility of our estimates presented depends on the assumption of random selection of households into migration conditional on the observables. If the decision to migrate is driven by unobservables, which also is correlated with possession of visible consumption goods, estimates of the impact might be biased. Accordingly, the first question we ask is whether estimates would change substantially and turn out statistically insignificant once we assume a greater extent of influence of unobservables?

To examine this, we employ the coefficient stability test first introduced by Altonji et al. (2005) and later extended by Oster (2019). The method provides a strategy to generate bounds of the treatment coefficient accounting for the unobservables. As discussed earlier, it assumes that the relationship between the treatment and the observables is informative about the relationship between the treatment and unobservables. The $\beta$ is calculated after accounting for the unobservables by assuming an arbitrary value of $R_{max} \epsilon [\tilde{R}, 1]$, where $\tilde{R}$ is the $R$-squared in the regression with controls. We use the value of $R_{max}$ suggested by Oster (2019), $R_{max} = 1.3 * \tilde{R}$. We calculate the value for the degree of selection on unobserved variables relative to that for observables, $\delta$ for which $\beta = 0$, for the given $R_{max}$. The value of $\delta$ (say $\delta = \hat{\delta}$) indicates that the unobservables have to be $\hat{\delta}$ times as important as the observables to get a treatment effect of zero. With an extensive set of controls, we can infer that the true $\beta$ would be significantly different from 0 if $\hat{\delta}$ exceeds the value one. However, if it remains below 1, it is more likely that the true $\beta$ would be statistically indistinguishable from zero in the presence of potential unobservables.

The value of $\hat{\delta}$, for each of the three controlled regressions are presented in the last row of Table 2. It is 3.06 for the regression of possession of luxury houses, which implies that the extent of unobservables has to be over 3 times that of the observables to bring the value of estimated impact to zero. As Oster (2019) indicates, this might be too high. Therefore, it can be safely assumed that international migration can significantly drive-up consumption of visible goods even after taking unobservables into consideration. Notably, for the other two outcome variables as well, we find the value of $\hat{\delta}$ to be 2.4 and 2.0, which is considerably greater than the threshold of 1.

*IV regressions*

Despite running the coefficient stability test, where we fail to reject the null hypothesis of no impact of transnational migration, we employ IV regression to ensure our estimates are robust. The IV we use, as mentioned, is the proportion of households having an international migrant within the *taluk* from the same religion as that of the corresponding household. The second stage regression results from the endogeneity corrected Two Stage Least Squares (2SLS) are presented in Panel A of Table 3 along with the F-statistics from the first stage. Notably, the F-statistic of our IVs remains consistently well above the thumb rule of 10.[9]

**Table 3**: IV regression to estimate the effect of transnational migration on visible consumption

| | (1) Luxury House | (2) 5 least possessed | (3) VCI |
|---|---|---|---|
| Panel A – 2SLS | | | |
| Migrant household | 1.174*** | 0.661** | 0.377** |
| | (0.319) | (0.296) | (0.178) |
| F-statistic | 274.8 | 274.8 | 274.8 |
| Marginal Effect | 0.376*** | 0.184** | |
| | (0.112) | (0.839) | |
| Panel B – Plausibly exogenous IV | | | |
| Migrant household | 0.582*** | 0.242*** | 0.643*** |
| | (0.114) | (0.087) | (0.178) |
| Migrant household (with uncertainty) | 0.582*** | 0.242*** | 0.643*** |
| | (0.116) | (0.088) | (0.181) |
| Controls | Yes | Yes | Yes |
| Observations | 14325 | 14325 | 14325 |

*Note. Estimations from IV probit regressions are presented in column (1) and (2) and for column (3), those from IV 2SLS regression are presented. The numbers in the parenthesis are the standard error clustered at the locality level. The F-statistic refers to the Wald version of the Kleibergen and Paap rk-statistic. The dependent variables are as follows (i) column (1): possession of luxury house, (ii) column (2): possession of at least one out of five least possessed goods in the locality and (iii) column (3): visible consumption index (VCI). The main variable of interest is whether the household has at least one international migrant. The IV is the proportion of households having an international migrant within the taluk from the same religion as that of the corresponding household. The first stage results are given in Table A4. The control variables are those listed in Table A2 and district fixed effect.* $*** \ p < 0.01$; $** \ p < 0.05$; $* \ p < 0.1$. *Two rows in Panel B give the value of the estimates assuming the IVs to be plausibly exogenous without and with uncertainty. The row 'Plausibly exogenous' assumes* $\Omega_r = 0$ *and 'with uncertainty' uses* $\Omega_\gamma = (0.125\sqrt{s_i^2 + s_r^2})$.

The findings indicate that having at least one international migrant in the household lead to a significant increase in likelihood of possession of a luxury house and at least one among the five least possessed goods within the locality and the VCI. We also find similar results using remittances received as the main variable of interest (see Table A7). This aligns with the naive probit regression estimates presented in Table 2. From the IV regressions, we also find an increase in the likelihood of possessing luxury house by 38 percentage points for households with at least one international migrant. The VCI is found to increase by about 0.4 standard deviations while the likelihood of possessing at least one out of five least possessed goods increases by 18 percentage points. So, to sum up, IV regressions also show a statistically significant and positive relationship between international migration and visible spending, thus indicating that the findings are robust to endogeneity concerns.

[9] Conventional thumb-rule for considering a t-ratio inferences is an F-statistic value more than 10. Lee et al. (2022) has shown that a true 5 percent test requires an F value more than 104.7. All of our baseline models are consistent with this stricter threshold as well.

### 4.2 Robustness

*Plausible Exogeneity of the IV*

The 2SLS approach assumes strict exogeneity of the IV, which implies that the relationship of the IV with the error terms determining the dependent variable would only manifest through the main variable of interest (international migration in our case) and not through any other channels. Conley et al. (2012) relax this assumption of the zero direct effect and allow to offer a prior value or a range of values for this relationship, $\gamma$. We use a zero-first-stage test on particular subsample, for which the indirect effect is zero, to estimate the exact value of $\gamma$, instead of choosing arbitrary prior values as recommended by van Kippersluis & Rietveld (2018). In our case, the sub-sample of non-migrants qualifies for estimating the direct effect of the IV on the dependent variable since the indirect impact through migration would be zero for them. The results of zero first-stage sample are provided in Table A5. The results of this exercise with the plausibly exogenous conditions are provided in Panel B of Table 3. The first row assumes no uncertainty over the $\hat{\gamma}$, whereas the other row allows for some uncertainty by allowing non-zero elements in the variance-covariance matrix, $\Omega_r$. Following van Kippersluis & Rietveld (2018), we set $\Omega_\gamma = (0.125\sqrt{s_i^2 + s_r^2})$, where $s_r$ is the standard error of $\hat{\gamma}$ of the non-migrant group of households and $s_i$ is the standard error with the migrant households. The treatment effect is found to be statistically significant in all the models in line with our baseline findings.

*IPWRA*

Next, we use the IPWRA method to address the endogeneity concerns of our treatment variable, which is whether a household currently has an international migrant. In the first step, IPWRA estimates a probit model considering the treatment status as the outcome variable, thus generating a propensity score of being treated for each observation. Let us denote the predicted probability as $\hat{P}(Treat_i = 1 \mid X_i)$, and these propensity scores are used to generate the Inverse Probability Weights (IPW). Therefore, for the treated group, $IPW_i = 1/\hat{P}(Treat_i = 1 \mid X_i)$ and for the control group, $IPW_i = 1/(1 - \hat{P}(Treat_i = 1 \mid X_i))$. In other words, IPWRA gives disproportionately higher weights to the observations that are less likely to be in their respective groups. In the second stage, the baseline outcome model is estimated using these weights in the regression. The estimates from IPWRA are doubly-robust, as it requires only one of the outcome and treatment models to be correctly specified (Imbens & Wooldridge, 2009; Wooldridge, 2007, 2010). Results from these regressions are presented in Table A6, and they indicate a qualitatively similar association of international migration with visible consumption.

*Alternate variable of interest*

In the paper, the primary variable of interest is whether the household has an international migrant or not. We now consider a different variant of this variable: the current amount of remittance received from the migrants by the sampled household every year. To avoid getting an undefined value for households without migrants while taking logarithmic value, we use the inverse hyperbolic sine transformation value (Aihounton & Henningsen, 2021; Burbidge et al., 1988). This ensures that the value of this variable would be zero for non-migrant households. The results from naïve OLS and Probit models, along with the IV regressions with the three outcome variables considered, are presented in Table A7. We observe our inferences to hold qualitatively- households receiving higher remittance from the international migrants are more likely to possess luxurious house and at least one among the least possessed assets within the locality apart from having a higher VCI on average.

### 4.3 Plausible mechanisms

This section discusses what drives this disproportionate spending on visible consumption among migrant households. What are the potential channels or reasons that explain the higher conspicuous spending among migrants? We empirically test two possible channels: taste-based changes in consumption pattern, and a peer or reference group effect. While the taste-based consumption pattern change channel reflects on the direct utility from the consumption of the visible goods, the reference group effect could mirror both a consumption pattern change through social learning and social utility (Bursztyn et al., 2014). We consider each of these channels separately.

***Taste-based change***

Migration of household members can result in changes in taste, influencing the consumption of visible durable goods, among others. Related literature has shown how changes in attitudes and utility function can be linked with migration which in turn can lead to favorable outcomes in terms of health and education. For example, Hildebrandt et al. (2005) point out how acquiring health knowledge influenced health outcomes back at home among the migrants from Mexico to the US. Because such knowledge acquisition or social learning can happen in our case as well since the migrants are exposed to a different lifestyle and consumption pattern in comparison to others in their hometown, and it could possibly lead to incremental spending on visible goods. To explore whether transnational migration is associated with a change in overall consumption pattern, we use three proxies of general consumption as outcome variables: total consumption, food, and educational expenditure. The results, presented in Table 4, indicate that households with a migrant have significantly higher total consumption expenditure at 10% level with no observed difference in food or educational expenditure. Since food and education expenditure are relatively non-visible in nature, an insignificant effect of migration on these expenditures possibly suggests that there has not been a general shift in taste and preferences. Nevertheless, it is possible that there is a preferential or taste-based change only for visible consumption goods among the migrants. Therefore, we cannot rule out the possibility of a migration-driven taste-based change in explaining higher status good consumption. However, the available data does not allow us to test this possibility and therefore studying the implications of this channel necessitates further research.

**Table 4**: Alternative Channels – Preference Change

| | (1)<br>Total Consumption Expenditure | (2)<br>Food Expenditure | (3)<br>Education Expenditure |
|---|---|---|---|
| Migrant Household | 0.314* | 0.021 | -0.799 |
| | (0.179) | (0.178) | (0.981) |
| Controls | Yes | Yes | Yes |
| F-statistic | 139.9 | 139.9 | 139.9 |
| Observations | 14325 | 14325 | 14325 |

*Note. Estimations from IV regressions are presented. The numbers in the parenthesis are the standard error clustered at the locality level. The F-statistic refers to the Wald version of the Kleibergen and Paap rk-statistic. The IV is the proportion of households having an international migrant within the taluk from the same religion as that of the corresponding household. All models have control variables that are listed in Table A2 and district fixed effect.*
$*** \ p < 0.01$; $** \ p < 0.05$; $* \ p < 0.1$.

***Reference Group Change***

Our second proposition is a reference group or peer effect that might drive up the visible spending among migrants. This is directly derived from the literature documenting the reference group income distribution effects on visible consumption (Chai et al., 2019; Charles et al., 2009; Roychowdhury, 2016). For example, Charles et al. (2009) show reference group income characteristics explain higher visible good consumption among Blacks and Hispanics in the US. Using a similar argument, we utilize different indicators that capture the distributional parameters of the income of the reference group. In our context, we consider that the reference group for a migrant household is other migrant households in the locality.

**Table 5**: Alternative Channels – Reference Group Change

| | Luxury House | | | | | |
|---|---|---|---|---|---|---|
| | (1) | (2) | (3) | (4) | (5) | (6) |
| Migrant Household | 1.174*** | 1.390*** | 1.496*** | 1.195*** | 1.292*** | 1.556*** |
| | (0.319) | (0.352) | (0.450) | (0.333) | (0.325) | (0.394) |
| Reference Group Income | | | | | | |
| Mean | | -0.194*** | | | | 0.046 |
| | | (0.072) | | | | (0.110) |
| Median | | | -0.711** | | | -1.018** |
| | | | (0.343) | | | (0.444) |
| 90th percentile | | | | -0.186 | | 0.214* |
| | | | | (0.147) | | (0.121) |
| Inequality | | | | | 1.416** | -0.106 |
| | | | | | (0.707) | (0.687) |
| F-statistic | 274.8 | 136.5 | 68.53 | 173.1 | 189.8 | 61.85 |
| Observations | 14325 | 14325 | 14325 | 14325 | 14325 | 14325 |

*Note. Estimations from IV regressions are presented. The numbers in the parenthesis are the standard error clustered at the locality level. The F-statistic refers to the Wald version of the Kleibergen and Paap rk-statistic. The main variable of interest is whether the household has at least one international migrant. The dependent variable is possession of luxury house. The results with other two dependent variables are given in Table A8. The IV is the proportion of households having an international migrant within the taluk from the same religion as that of the corresponding household. All models have control variables that are listed in Table A2 and district fixed effect.* $^{***}p < 0.01$; $^{**}p < 0.05$; $^{*}p < 0.1$.

Similarly, non-migrant households in the locality serve as the reference group for a non-migrant household. For this, we calculate the locality-wise mean, median, and 90th percentile income levels along with another indicator of inequality, which is measured as the 90th percentile income level divided by the 25th percentile for migrant households as well as the non-migrants. Table 5 presents the results from the regression of luxury house possession with the above-mentioned income distribution parameter values of the reference group at the locality level as additional controls. The first column reports the estimate from our baseline model. We observe no changes in the effect size when we control for the reference group income level and inequality. This suggests that unlike Charles et al. (2009) where the reference group income level and variation explain the entire difference in visible consumption between races, here it may explain the migration effect that we observe but to a limited extent.[10] Interestingly, from the qualitative

[10] Table A8 presents the reference group income-controlled regression results for the other dependent variables. We also implement causal mediation analysis in the spirit of Imai et al. (2011) and Carpena & Zia (2020) to check how much of the relationship between migration and visible consumption is mediated through the variables we used to capture preference change and reference group income change. The results from mediation analysis are provided in Table A9. Only a small portion of the relationship is mediated through these channels, 17% being the maximum. We

surveys we conducted, the details of which is given in the next section, one of the respondents reported the following when asked about the reason for migrants possessing a luxurious house: "*The migrants see all the other migrants doing so, so they also indulge in such consumptions to keep up with them*". This indicates a possibility of reference group effect driving conspicuous spending among migrants. However, our results show a limited role of the channel in explaining the migration effect.

***Information gap***

Having discussed two channels that can potentially explain the differences in visible spending between migrants and non-migrants, we introduce a third channel, information gap about one's income as a potential explanation for the observed effects. We propose that the differences in incentives to signal status due to this information gap[11] difference between migrants and non-migrants could be a probable mechanism. The argument is that the importance of social signaling, wherein an individual derives utility from status, depends on the beliefs and information others have about one's income/wealth. Notably, information within the community on the income levels of the migrant is often disproportionately more incomplete. This becomes clear from a set of literature that documents the existence of high information gap between the migrant and family or society back home (Ambler, 2015; Chen, 2013; de Laat, 2014; McKenzie et al., 2013). Seshan & Zubrickas (2017), for example, has shown the existence of the same phenomenon for the case of Kerala migration within the family. If the income levels of migrants are unknown to the family, it is very less likely that others within the community do have much information on their actual income as well. This can be leveraged by the migrant through higher spending in status goods to showcase the income level and derive utility through higher social status. The problem of this information gap within the community for non-migrants can also be prevalent. Still, the extent would possibly be higher for migrants, whose work profile and living standards often remains unobservable. Accordingly, the migrants can then move up the social status ladder by potentially spending more on visible goods. Arguably, they find this conducive to do in comparison to the non-migrants because of larger information gap about their income and job profile within society.

As we expect all migrant households have same level of information gap in society after controlling for their background characteristics, the data does not allow us to test the information gap channel explicitly in the case of migrants. To test the significance of this mechanism, therefore, we restrict our sample to non-migrant households and generate an index of information gap using the second

---

only present the results of total consumption expenditure and median income, but all the variables provide similar results with no unreported variable mediating more than 12%.

[11] For the purposes of this paper, we use the term "information gap" to refer to the extent of imperfection in society's information about the income/ wealth level of individuals/households. Information gap can be seen as the magnitude of information asymmetry on actual income/ wealth level.

moment of job specific earnings. We calculate the standard deviation of earnings from each job category across the entire sample and assign the highest standard deviation within the various jobs held by household members. The idea is that the income from jobs with higher variability in earnings will be less predictable, and thus increases the information gap in society about the actual earnings. We run OLS regression along with coefficient stability test to assess whether households with increased income unpredictability have higher visible consumption. The results, presented in Table 6, is indicative of a significantly positive association between the two.[12]

**Table 6**: Effect of Income Unpredictability on Visible Consumption

| | (1)<br>Luxury House | (2)<br>5 least possessed | (3)<br>VCI |
|---|---|---|---|
| Household Income<br>Unpredictability | 0.284***<br>(0.028) | 0.408***<br>(0.025) | 0.289***<br>(0.017) |
| Controls | Yes | Yes | Yes |
| Observations | 10480 | 10480 | 10480 |
| $\hat{\delta}\ for\ \beta = 0$ | 1.541 | 1.676 | 1.598 |

*Note. Estimations from probit regression are presented in columns (1) and (2) and for column (3), those from OLS regression are presented. The numbers in the parenthesis are the standard error clustered at the locality level. The dependent variables are as follows (i) column (1): possession of luxury house, (ii) column (2): possession of at least one out of five least possessed goods in the locality and (iii) column (3): visible consumption index (VCI). The main variable of interest, Household Income Unpredictability, is the hyperbolic sine of the earnings standard deviation of the income source of household. The control variables are those listed in Table A2 and district fixed effects.* $*** \ p < 0.01$; $** \ p < 0.05$; $* \ p < 0.1$. *In the last row, the value of* $\hat{\delta}$*, from the coefficient stability (Oster) test, for which the coefficient estimate turns zero is given.*

Even though we find a positive association between the information gap and visible consumption, we cannot directly test how much our baseline finding is mediated through this channel. However, given the fact that other channels are unable to explain much of the migration-visible consumption relationship, the information gap channel is likely to be a significant contributor to the relationship. To explore further on this, we conduct a qualitative exploration and then use a theoretical framework to explain the mechanism of increased incentives through the information gap in the following section.

## 5 Conspicuous spending and information gap

We put forward higher information gap on the job and income of the migrant as one of factors driving the migration-visible consumption relationship. To understand the prevalence of status signaling among migrants and the underlying reasons, we conduct interviews with 20 international migrants and return migrants from two districts of Kerala, Malappuram and Trissur, following convenience sampling. The survey was conducted between June and August 2022. The survey used a semi-structured interview to gather information on the social considerations of consumption among migrant households. The interviews were conducted in the local language to ensure that participants were comfortable and could express themselves freely. The sample had 10 individuals from each district with in total 17 males and 3 females. The age of the respondents ranged from 28 to 60 years old. 19 participants were married and had children. 13 respondents had only done some schooling and the remaining held a degree. The number of return migrants were four. The questions focused on the reasons behind the higher level of consumption of visible goods among migrants and the information asymmetry on migrants' earnings. In the interview, one international return migrant shared his thought about possession of luxurious houses and the changing nature of status signaling.

[12] We also define the household income predictability by generating dummy variables from the standard deviation distribution; taking value one if the standard deviation is higher than the median and mean of the distribution respectively. The results are similar to what we find with the inverse hyperbolic sine transformation of standard deviation. They are presented in Table A10.

"*The status symbols have changed over the time. In earlier days, the height of the heap of straws in the backyard after harvest was considered to be the symbol of status. Nowadays, it's the height of the gate of the house and people are competing each other*"

In Kerala, the information asymmetry in migrants' earnings has been already documented by Seshan & Zubrickas (2017), and our survey also elicited the existence of information asymmetry. All of our survey respondents in fact reported of this information gap. They believe that neighbors and others back in their village often do not know about the exact occupation when they migrate. Accordingly, there is uncertainty about the earnings as well. One respondent reported that "*the information gap is so wide that we could often find people doing jobs that they might not do back home due to status concerns, like working in fish markets.*" More than half of the individuals interviewed informed us that families do not have any idea about the earnings when their family members go abroad for working. This social feature is also well documented in popular media and fictions. For instance, the following is a famous monologue from the Malayalam movie "*Pathemari*", which narrates the lives of gulf migrants from Kerala.

*"When we think we're toiling for the sake of our loved ones we don't feel the fatigue. We feel happy when the money sent would have fulfilled a need. Most of us are here only physically. Even my wife is unaware of what my job is or my salary, not because I'm ashamed, but not to let them know about the hardships. When we send Rs. 10000, what do people think? Out of 20000 we earn, we sent 10000. But in reality, we earn 7000 and take a loan for the rest."*

In this situation, it is possible that the prevalent information asymmetry is leveraged by the migrants to signal their earnings in their villages. One potential way of doing so is spending on luxurious houses and other items which gives a visible signal about the income levels of the migrants and hence status within the community. As the information gap increases, in our case through migration, the incentives to signal status through consumption increases and visible consumption is expected to increase.

***Theory***

We formalize the information gap mechanism by developing a simple model of conspicuous consumption under asymmetric information. Our theoretical framework closely follows Corneo & Jeanne (1997). However, unlike in Corneo & Jeanne (1997), in the present analysis there are two groups of consumers, and these two groups differ in terms of observing their income by others in society.

Consider a society having a continuum of households (consumers) of mass one. Households are of two types: (a) migrant ($M$), whose primary source of income is the remittances sent by international migrant(s) from the household and (b) villager ($V$), who works in a nearby place and does not receive any remittances. In this society $\lambda \in (0,1)$ the proportion of households are of $M$-type and $(1-\lambda)$ proportion are of $V$-type. Households are indexed by $r \in [0,1]$. The index $r$ represents the household's rank in the income order: $y_r > y_{r'} \Leftrightarrow r < r'$, $\forall\, r, r' \in [0,1]$, where $y_r$ and $y_{r'}$ denote incomes of households having ranks $r$ and $r'$, respectively. That is, a lower rank indicates a better position in the income hierarchy.

$M$-type households' income is distributed according to the continuous density function $f(y) > 0$ with support $[\underline{y}^M, \overline{y}^M]$, $0 < \underline{y}^M < \overline{y}^M < \infty$, where the corresponding cumulative distribution function is given by $F(y)$. The density function and cumulative distribution function of $V$-type households' income are given by $g(y) > 0$ and $G(y)$, respectively, which are continuous and defined over $[\underline{y}^V, \overline{y}^V]$, $0 < \underline{y}^V < \overline{y}^V < \infty$. Income distribution of all households, both $M$-type and $V$-type households, together is assumed to be given by a continuous density function over the support $[\min\{\underline{y}^M, \underline{y}^V\}, \max\{\overline{y}^M, \overline{y}^V\}]$. Income distributions are exogenously given and are common knowledge to all. However, while income of each $V$-type household is known to all, a $M$-type household's income is her private information. That is, there is perfect information regarding each $V$-type household's income and, hence, their ranks. In contrast, only a $M$-type household knows her true income and rank in the income hierarchy, not observable by others in society. Households are assumed to be concerned about their relative status in society, which depends on their ranks in the income hierarchy.

There are two consumption goods, good 1 and good 2. Good 1 is a normal good and a household's consumption of this good is not observable to others. On the other hand, the public observes households' consumption of good 2, which is a conspicuous good and is referred as the *status good*. The price of good 2 is $p$ ($> 0$) and the price of good 1 is normalized to be equal to one, i.e., good 1 is the numeraire good. We assume that the status good is indivisible, and households do not buy more than one unit of the status good.

Unlike the numeraire good, the status good does not yield any intrinsic utility a la Veblen (1899). Consumption of the status good by a household may send information regarding that household's income rank to the public, which is the only purpose of consumption of status good. Let $x_r$ denote the quantity of good 1 consumed by the $r$-th ranked household and $\delta_r$ be the indicator variable that takes value 1, if the $r$-th ranked household buys the status good, and 0, if she does not. Note that, for each $V$-type household, the public observes both her rank and consumption of the status good, if any. In contrast, for each $M$-type household, the public observes only her consumption of the status good, if any, and not her rank in the income hierarchy.

Households' preferences are assumed to be identical and represented by the following utility function, as in Corneo & Jeanne (1997).

$$U(x_r, \delta_r) = u(x_r) + v(\delta_r), \qquad (3)$$

where $u(x_r)$ is the utility of the numeraire good and $v(\delta_r)$ is the utility of the status good. $u(\cdot)$ is a twice continuously differentiable, strictly increasing and concave function; $u'(0) = \infty$ and $\lim_{c \to +\infty} u'(c) = 0$. As discussed before, consumption of the status goods sends information regarding the income rank of the household, the utility of the status good is derived through its possible effect on others' beliefs regarding the household's rank in the income hierarchy. All households have access to the same information: (a) income level of each $V$-type household, (b) status good consumption by each household and (c) distribution of income of each type of households. Thus, status good consumption by a household carries the same signal to all others and everyone updates their beliefs in the same way. We assume that $v(\delta_r) = E(a(r)|\delta_r)$, where $a(r)$ is the rank-utility of the household having rank $r$ and $E(a(r)|\delta_r)$ is the expected value of $a(r)$ given consumption of status good $\delta_r \in \{0, 1\}$. The rank-utility function $a(r)$ is assumed to be continuous and strictly decreasing in $r$: $a(r): [0, 1] \to [\underline{a}, \overline{a}]$; $a(1) = \underline{a}$ and $a(0) = \overline{a}$ are finite. That is, better rank (a lower value of $r$) in the income hierarchy brings higher social status and, thus, yields higher utility. Therefore, we can rewrite the utility function in (3) as follows.

$$U(x_r, \delta_r) = u(x_r) + E(a(r)|\delta_r) \qquad (4)$$

Households maximize their respective utility subject the following budget constraint.

$$x_r + p\,\delta_r \leq y_r \qquad (5)$$

Now, first consider $V$-type households. Since each $V$-type's income and, hence, rank is observable by all, we have $(a(r)|\delta_r) = a(r)\ \forall\ \delta_r \in \{0,1\}$. That is, $V$-type households do not get any benefit from consuming the status good. Hence, it is optimal for each $V$-type household not to consume the status good ($\delta_r = 0$).

Next, consider $M$-type households. Maximum utility levels of a $M$-type household (a) if she consumes the status good ($\delta_r = 1$) and (b) if she does not consume the status good ($\delta_r = 0$) are, respectively, given by $U^M_{\delta_r=1} = u(y_r - p) + E(a(r)|\delta_r = 1)$ and $U^M_{\delta_r=0} = u(y_r) + E(a(r)|\delta_r = 0)$. Therefore, it is optimal for a $M$-type household to consume the status good, if the following condition holds true.

$$U^M_{\delta_r=1} \geq U^M_{\delta_r=0} \Leftrightarrow E(a(r)|\delta_r = 1) - E(a(r)|\delta_r = 0) \geq u(y_r) - u(y_r - p) \quad (6)$$

That is, a $M$-type household consumes the status good, if its signal value $s = [E(a(r)|\delta_r = 1) - E(a(r)|\delta_r = 0)]$ is greater than the threshold level $[u(y_r) - u(y_r - p)]$. Assuming that $s > 0$, condition (6) is satisfied if and only if $y_r \geq y_o(s,p)$, where $y_o(s,p)$ is the threshold level of income given by (7) (Corneo & Jeanne, 1997).

$$\left.\begin{array}{ll} y_o(s,p) = p, & \text{if } s \geq u(p) - u(0)\text{; and} \\ s = u\big(y_o(s,p)\big) - u(y_o(s,p) - p), & \text{if } s \leq u(p) - u(0) \end{array}\right\} \quad (7)$$

That is, $M$-type households are divided into two groups: one who are sufficiently rich and consumes the status good, and the other who are poor and does not buy the status good. The number of $M$-type households who buy the status good is given by $m = [1 - F(y_o(s,p))]\lambda$, which is also the demand for the status good.

Note that the number of $V$-type households whose income is greater than $y_o(s,p)$ is given by $n = [1 - G(y_o(s,p))](1-\lambda)$. Therefore, in society there are total $n + m = [1 - G(y_o(s,p))](1-\lambda) + \big[1 - F\big(y_o(s,p)\big)\big]\lambda$ households are rich ($y \geq y_o(s,p)$), and remaining $(1 - n - m)$ households are poor. Owners of the status good send the signal that their rank in income hierarchy is better than $n + m$. The value of such a signal can be expressed in terms of the difference between the average rank utility of $n + m$ rich households and of $1 - n - m$ poor households, as follows.

$$s = \frac{\int_0^{n+m} a(r)dr}{n+m} - \frac{\int_{n+m}^1 a(r)dr}{1-n-m} = \frac{\int_0^{n+m}[a(r) - \bar{a}]\, dr}{(n+m)(1-n-m)} = \sigma(n+m), \quad \text{say,} \quad (8)$$

where $\overline{a} = \int_0^1 a(r)dr$ is the average rank utility over all the households, M-types as well as V-types, in the society. Since $a(r)$ is strictly decreasing in $r \in [0, 1]$, $\int_0^{n+m} [a(r) - \overline{a}]\, dr > 0 \Rightarrow s > 0$. That is, the signal value of the status good is always positive to rich $M$-type households.

From the above analysis we can state the following.

**Proposition 1:** *Suppose that (i) a $M$-type household's income is her private information, while a $V$-type household's income is common knowledge, and (ii) income distribution of each type of households and the joint income distribution of all households taken together are continuous and are common knowledge. Then, in the equilibrium relatively rich $M$-type households consume the status good, while none of $V$-type households consume the status good.*

From Proposition 1 it follows that, in the equilibrium under asymmetric information, a migrant household is more likely to spend on status goods than a non-migrant household, ceteris paribus. Now, from (8) it follows that the signal value $s$ of the status good consumption depends on the total number of rich households (all those whose income $y_r$ is greater than $y_o(s, p)$) in society: $s = \sigma(n + m)$. It is easy to check if the rank utility function $a(r)$ is quadratic and convex (concave) in $r \in [0, 1]$, then $\sigma'(r) <$ ($>$) 0.[13] When the signal value of the status good decreases (increases) with an increase in the number of its consumers, the status good consumption is said to be snobbish (conformist). In the present context, when the rank utility function is convex (concave), the signal value of the status good decreases (increases) with an increase in the number of rich $V$-type households ($n$), for any given number of the status good consumers among $M$-type households ($m$), and vice-versa.

First, suppose that the rank utility function is convex. Then, the gain in social status by climbing up one position in the income hierarchy is more for a higher ranked household compared to that of a lower ranked household. As more households are identified with the rich the signal value of the status good falls. In other words, a $M$-type household derives higher gain is social status by consuming the status good, when fewer households can identify with the rich. That is, when existing social norm is such that the rank utility function is convex, there will be snobbish behavior in status good consumption by the $M$-type. Clearly, in such a scenario, if the income distribution of the $V$-type households is more positively skewed, the number of rich $V$-type households ($n$) is likely to be greater given the proportion of $M$-type households in the population ($\lambda$), and, hence that would reduce the signal value of the status good and a fewer number of $M$-type households will consume the status good (snob effect).

[13] Differentiating both sides of $\frac{\int_0^{n+m}[a(r)-\overline{a}]dr}{(n+m)(1-n-m)} = \sigma(n+m)$ with respect to $(n + m)$ and rearranging terms we get, $a(n + m) = \overline{a} + [1 - 2(n + m)]\sigma(n + m) + (n + m)[1 - (n + m)]\sigma'(n + m)$. Clearly, if the rank utility function $a(r)$ is quadratic in $r$, then (i) $\sigma(r)$ is linear in $r$ and (ii) $a''(r) = -6\,\sigma'(r)$, $\forall\, r \in [0, 1]$. If the rank utility function is linear, the signal value of the status good does not depend on $n + m$.

Next, suppose that the existing social norm is such that the loss in social status (i.e., decrease in rank utility) due to losing one position in the income hierarchy is more for a lower ranked households compared to that of a higher ranked household (which implies that the rank utility function is concave). Then, the loss in social status for being identified with the poor is higher and, thus, an increase either in the number of $M$-type households consuming the status good or/and in the number of rich $V$-type households leads to a more than proportionate decrease in the social status of $M$-type households who do not consume the status good. In such a scenario, the signal value of the status good increases with the number of $M$-type status good consuming households, for any given income distribution of the $V$-type households. In other words, the existing social norm summarized by concave rank utility function leads to conformist behavior among the $M$-type households. It also follows that in such a society, if the distribution of the $V$-type households' income is more positively skewed, a greater number of $M$-type households will consume the status good.[14]

Further note that, we have $s = \sigma(n + m)$ and $n + m = [1 - G(y_o(s,p))](1 - \lambda) + [1 - F(y_o(s,p))]\lambda$, where $y_o(s,p)$ is given by (7). Therefore, we can write the following.

$$\frac{ds}{d\lambda} = \frac{\frac{\partial s}{\partial(n+m)}\frac{\partial(n+m)}{\partial\lambda}}{1 - \frac{\partial s}{\partial(n+m)}\frac{\partial(n+m)}{\partial y_o(s,p)}\frac{\partial y_o(s,p)}{\partial s}} \tag{9}$$

Now, (i) $\frac{\partial s}{\partial(n+m)} < (>)\ 0$, if $a''(\cdot) > (<) 0$;

(ii) $\frac{\partial(n+m)}{\partial\lambda} = G(y_0(s,p)) - F(y_0(s,p)) < (>)\ 0$, if $G(\cdot) < (>) F(\cdot)$;

(iii) $\frac{\partial(n+m)}{\partial y_o(s,p)} = -g(y_o(s,p))(1-\lambda) - f(y_o(s,p))\lambda < 0$; and

(iv) $\frac{\partial y_o(s,p)}{\partial s} = \begin{cases} \frac{1}{u'(y_o) - u'(y_o - p)} < 0, & if\ s \le u(p) - u(0) \\ 0, & if\ s \ge u(p) - u(0) \end{cases}$ (from (7)).

Therefore, we have the following.

---

[14] Since $\sigma'(m+n) > 0$ and $n = [1 - G(y_o(s,p))](1-\lambda)$ is higher whenever $[1 - G(y_o(s,p))]$ is greater.

Suppose that $a''(\cdot) > 0$, i.e., the existing social norm induces snobbish behaviour in status good consumption. Then, the following is true.

$\frac{ds}{d\lambda} < 0$, if $F(\cdot)$ first order stochastically dominates $G(\cdot)$. That is, a greater proportion of $M$-type households in society results in a decrease in the signal value of the status good, if the $M$-type's income distribution first order stochastically dominates the $V$-type's income distribution.

$\frac{ds}{d\lambda} > 0$, if $G(\cdot)$ first order stochastically dominates $F(\cdot)$. That is, the signal value of the status good increases with the increase in the proportion of $M$-type households in society, if the $V$-type's income distribution first order stochastically dominates the $M$-type's income distribution.

Suppose that $a''(\cdot) < 0$; i.e., the existing social norm induces conformist behaviour in status good consumption. Then, the following is true.

$\frac{ds}{d\lambda} > 0$, if $F(\cdot)$ first order stochastically dominates $G(\cdot)$ and $\frac{ds}{d(n+m)}\frac{\partial(n+m)}{\partial y_o(s,p)}\frac{\partial y_o(s,p)}{\partial s} < 1$, which always holds true if $s \geq u(p) - u(0)$. That is, under certain conditions, a greater proportion of $M$-type households in society results in an increase in the signal value of the status good, if the $M$-type's income distribution first order stochastically dominates the $V$-type's income distribution.

$\frac{ds}{d\lambda} < 0$ if $G(\cdot)$ first order stochastically dominates $F(\cdot)$ and $\frac{ds}{d(n+m)}\frac{\partial(n+m)}{\partial y_o(s,p)}\frac{\partial y_o(s,p)}{\partial s} < 1$, which always holds true if $s \geq u(p) - u(0)$. It implies that, under certain conditions, an increase in the proportion of $M$-type households in society reduces in the signal value of the status good, if the $V$-type's income distribution first order stochastically dominates the $M$-type's income distribution.

**Proposition 2:** *Suppose that (i) a $M$-type household's income is her private information, while a $V$-type household's income is common knowledge, and (ii) income distribution of each type of households and the joint income distribution of all households taken together are continuous and are common knowledge. Then, whether the signal value of the status good is increasing in the proportion of $M$-type households in society, or not, that depends on (a) the underlying social norm – conformist or snobbish and (b) relative income distributions of the two types of households.*

In this analysis we have assumed that households do not derive any intrinsic value by consuming the status good, which only helps to signal the income rank. Further, we assume that there is only one status good. However, there are multiple status goods and some of these may have both intrinsic value of consumption as well as the signal value. In such scenarios, Proposition 1 implies that, because of existing informational asymmetry regarding $M$-type households' rank in the income hierarchy, $M$-type households are more like to spend on consumption of status goods compared to $V$-type households, as observed in our empirical analysis.

**Table 7**: Theory Predictions: Testing for Snobbish or Conformist Behavior

| | All Sample | Higher Rich Migrants | Higher Non-Rich Migrants |
|---|---|---|---|
| | (1) | (2) | (3) |
| Migrant Household | 0.177*** | 0.201*** | 0.590*** |
| | (0.021) | (0.028) | (0.175) |
| Proportion of Migrants | 0.127 | 0.136 | 1.032*** |
| | (0.079) | (0.087) | (0.510) |
| Migrant Household * Proportion of Migrants | | -0.296*** | -0.714 |
| | | (0.099) | (0.640) |
| Proportion of rich non-migrants | 0.278*** | | |
| | (0.056) | | |
| Migrant Household * Proportion of rich non-migrants | -0.237*** | 0.089*** | 0.171*** |
| | (0.085) | (0.016) | (0.026) |
| Controls | Yes | Yes | Yes |
| Observations | 14325 | 9780 | 4545 |

*Note. Estimations from probit regression are presented. The numbers in the parenthesis are the standard error clustered at the locality level. The dependent variable in all estimations is possession of luxury house. Migrant household is the households that has at least one international migrant. All models have control variables that are listed in Table A2 and district fixed effect.* $*** \ p < 0.01$; $** \ p < 0.05$; $* \ p < 0.1$.

***Testing the Model Predictions***

Our data allows us to give two sets of suggestive evidence on whether the migrants exhibit snobbish or conformist behavior while spending on status goods. First, we deduced that, when status goods consumption is snobbish, "if the income distribution of the $V$-type households is more positively skewed, the number of rich $V$-type households ($n$) is likely to be greater given the proportion of $M$-type households in the population ($\lambda$), and, hence that would reduce the signal value of the status good and a fewer number of $M$-type households will consume the status good (snob effect)." Based on this, we interact the international migrant dummy with the proportion of rich non-migrants within the locality and regress it along with the individual variables, the set of control variables and the proportion of migrant households in the locality. Here, INR 7000, which is the 75$^{th}$ percentile per-capita monthly income is taken as the threshold beyond which the household is categorized as rich. We observe that the chances of migrants possessing a luxurious house fall disproportionately as the proportion of rich non-migrants rises in the locality (see Table 7). Therefore, the migrant households potentially exhibit snobbish behavior while signaling status.

Second, we argue above that "a greater proportion of $M$-type households in society results in a decrease in the signal value of the status good if the $M$-type's income distribution first-order stochastically dominates the $V$-type's income distribution" indicates snobbish behavior in status goods consumption. To test this, we interact the migrant dummy variable with the proportion of migrant households within the locality and regress it on status good possession separately for two types of localities: one with higher proportion of richer migrants and those with a higher proportion of richer non-migrants. We find that in localities from the former type, the likelihood of status good possession disproportionately reduces for migrant households as the proportion of migrants from the locality increases. In the other type of localities, (the one with a higher proportion of richer non-migrants), this relationship remains statistically insignificant, implying that migrant spending on status goods remain unchanged even when proportion of migrants around them increases (see Table 7). Hence, these two sets of evidence suggestively indicate that the migrants, while spending on status goods, follow a snobbish behavior and not a conformist one.

## 6 Concluding Remarks

Signalling of social status within the community through expenditure on visible goods or conspicuous consumption is documented to be among the common ways by which individuals establish themselves in the social hierarchy. While existing literature has indicated how status-seeking behavior varies across individuals and households from different races, social and income groups, focus on how migrants behave in this respect has received relatively less attention. Potential changes in the consumption preference, change in reference group, and increased information gap in their permanent residential localities can have differential implications on migrants' visible goods consumption patterns. In this context, using micro-level representative data from the Indian state of Kerala, this paper examines the effect of transnational migration on social signaling through expenditure on visible durable goods.

Using a variety of regression models and robustness checks that control for the potential unobserved heterogeneity, we find that international migrants are more likely to spend on visible durable goods. While we are unable to rule out the possibility of these results being driven by a consumption pattern change among the migrants and reference group income distribution, their role appears to be limited. In this context, we propose another mechanism of change: we argue that the utility derived from enhancing social status can potentially be higher among migrants because of higher information gap about their income in their residential localities. We utilize literature and qualitative interviews to justify our point. Next, we provide a theoretical framework to argue migrants often leverage this higher information gap to spend more on visible consumption goods that signal status in the community. Further, we derive, to test here, conditions to understand when status good consumption exhibits snobbish behaviour and when it corresponds to conformist behaviour. In our context, we document predominance of a snobbish behaviour through visible consumption among the migrants.

The information gap channel of conspicuous consumption, we introduce in this paper to throw light on the migration effect, has wider implications. The key takeaway is that information gap heterogeneity can lead to different levels of conspicuous consumption. In other words, higher information gap within the community can potentially incentivize individuals to undertake more conspicuous consumption to enhance their social position. Accordingly, it can explain the rise in middle-class consumerism, household debt, and debt financing in recent decades (Bartscher et al., 2020; Mian et al., 2020). The macro implications of such debt financing are well-documented in the literature; expansion of household debts is found to be interrelated with business cycles, and it predicts financial crises and a future reduction in aggregate demand and economic growth (Illing et al., 2018; Jordà et al., 2016; Kumhof et al., 2015; Mian et al., 2017, 2021; Mian & Sufi, 2018; Schularick & Taylor, 2012). Thus, the increase in the information gap in society could have broader effects on the economy.

*Appendix*

**A1 Migrant Questionnaire**

<table>
<tr><td colspan="2">Name:</td></tr>
<tr><td>Age:</td><td>Gender: Female/Male/Other</td></tr>
<tr><td>Marital Status:<br><br>Not Married/Married/Divorced/Widowed</td><td>Number of Children:</td></tr>
<tr><td colspan="2">Highest Education: No schooling/Primary/High School/Higher Secondary/Degree and above</td></tr>
<tr><td>Village:</td><td>District:</td></tr>
<tr><td colspan="2">Migration Status: Current Migrant/Return Migrant</td></tr>
<tr><td colspan="2">When you make big purchasing decisions, what are the general considerations you would have? (Probe about whether they consider what other people might think about the good they are purchasing)</td></tr>
<tr><td colspan="2">Do you feel any pressure to display your wealth through your purchases? If so, where do you think that pressure comes from?</td></tr>
<tr><td colspan="2">When we looked at the data, we found that migrants are likelier to have big houses than non-migrants. What do you think might be the contributing factors?</td></tr>
<tr><td colspan="2">Do you think people back in the village know much about the out-migrant's jobs and earnings?</td></tr>
<tr><td colspan="2">Other observations</td></tr>
</table>

**Table A1**: Difference between Migrant and Non-migrant Households (Key Variables)

| | Non-migrant HH | Migrant HH | Difference | S.E. |
|---|---|---|---|---|
| ***Main Dependent Variables*** | | | | |
| Luxury House | 0.39 | 0.61 | -0.22*** | (0.01) |
| Conspicuous Consumption HH | 0.32 | 0.51 | -0.19*** | (0.01) |
| Visible Consumption Index | -0.08 | 0.37 | -0.45*** | (0.02) |
| Total Remittances | 0.01 | 11.45 | -11.44*** | (0.03) |
| ***Control Variables*** | | | | |
| Total HH Income (Monthly) | 22769.29 | 30348.10 | -7578.81*** | (1239.67) |
| *Religion* | | | | |
| Hindu | 0.60 | 0.36 | 0.24*** | (0.01) |
| Christian | 0.21 | 0.20 | 0.01 | (0.01) |
| Muslim | 0.19 | 0.44 | -0.25*** | (0.01) |
| Rural Household | 0.52 | 0.52 | 0.00 | (0.01) |
| Male Household Head | 0.76 | 0.52 | 0.23*** | (0.01) |
| Household Size | 4.06 | 4.10 | -0.04 | (0.04) |
| Number of earning individuals in the HH | 1.41 | 1.94 | -0.53*** | (0.02) |
| *Ration Card Type* | | | | |
| Yellow or Pink (Below Poverty Line)# | 0.42 | 0.25 | 0.17*** | (0.01) |
| Blue (Above Poverty Line) | 0.37 | 0.52 | -0.15*** | (0.01) |
| White (Non-priority) | 0.21 | 0.23 | -0.02** | (0.01) |
| *Highest Education in the HH* | | | | |
| School or less | 0.29 | 0.10 | 0.19*** | (0.01) |
| Higher Secondary | 0.29 | 0.31 | -0.03** | (0.01) |
| College | 0.23 | 0.27 | -0.04*** | (0.01) |
| Higher Education | 0.19 | 0.32 | -0.13*** | (0.01) |
| ***Other Variables*** | | | | |
| Food Expenditure (Monthly) | 5766.70 | 6745.65 | -978.96*** | (117.26) |
| Education Expenditure (Yearly) | 15345.14 | 22589.36 | -7244.23*** | (2531.32) |
| Total Consumption Expenditure (Monthly) | 13720.92 | 17691.90 | -3970.97*** | (354.07) |
| No. of Observations (Total – 14325) | 11815 | 2510 | | |

# The group also includes Most Economically Backward (Yellow) card holders. $*** \ p < 0.01; ** \ p < 0.05; * \ p < 0.1$

**Table A2**: Effect of transnational migration on visible consumption – Full Controls

| | (1)<br>Luxury House | (2)<br>5 least possessed | (3)<br>VCI |
|---|---|---|---|
| Migrant Household | 0.433*** | 0.358*** | 0.228*** |
| | (0.040) | (0.039) | (0.023) |
| Household Income | 0.041*** | 0.041*** | 0.045*** |
| | (0.008) | (0.010) | (0.006) |
| Religion<br>(Reference – Hindu) | | | |
| Christian | 0.154*** | 0.224*** | 0.190*** |
| | (0.047) | (0.040) | (0.024) |
| Muslim | 0.284*** | 0.229*** | 0.215*** |
| | (0.042) | (0.043) | (0.025) |
| Rural Household | -0.228*** | -0.154*** | -0.171*** |
| | (0.054) | (0.047) | (0.026) |
| Male Household Head | 0.028 | 0.058** | 0.038*** |
| | (0.030) | (0.029) | (0.014) |
| Household Size | 0.005 | 0.013* | 0.008* |
| | (0.008) | (0.008) | (0.005) |
| No. of earning individuals | -0.076*** | -0.088*** | -0.012 |
| | (0.017) | (0.016) | (0.009) |
| Ration Card Type<br>(Reference – Yellow/Pink) | | | |
| Blue | 0.626*** | 0.657*** | 0.391*** |
| | (0.032) | (0.036) | (0.016) |
| White | 1.182*** | 1.360*** | 1.012*** |
| | (0.043) | (0.044) | (0.026) |
| Highest Education in Household<br>(Reference – School or less) | | | |
| Higher Secondary | 0.271*** | 0.285*** | 0.182*** |
| | (0.035) | (0.039) | (0.016) |
| College | 0.467*** | 0.593*** | 0.386*** |
| | (0.040) | (0.041) | (0.019) |
| Higher Education | 0.765*** | 1.056*** | 0.774*** |
| | (0.043) | (0.049) | (0.028) |
| District FE | Yes | Yes | Yes |
| Observations | 14325 | 14325 | 14325 |

*Note. Estimations from probit regression are presented in columns (1) and (2) and for column (3), those from OLS regression are presented. The numbers in the parenthesis are the standard error clustered at the locality level. The dependent variables are as follows (i) column (1): possession of luxury house, (ii) column (2): possession of at least one out of five least possessed goods in the locality and (iii) column (3): visible consumption index (VCI).* $*** \ p < 0.01$; $** \ p < 0.05$; $* \ p < 0.1$.

**Table A3**: Coefficient Stability Conditions

| | Baseline Effect (Std. Error) [R2] | Controlled Effect (Std. Error) [R2] | Identified Set | $\hat{\delta}\ for\ \beta = 0$ |
|---|---|---|---|---|
| Luxury House | 0. 215*** (0.015) [0.027] | 0.145*** (0.013) [0.244] | [0.111, 0.145] † | 3.061 |
| 5 least possessed | 0. 195*** (0.013) [0.024] | 0.111*** (0.012) [0.287] | [0.072, 0.111] † | 2.384 |
| VCI | 0. 447*** (0.029) [0.029] | 0.228*** (0.023) [0.404] | [0.130, 0.228] † | 2.040 |

*Note. The treatment variable is 'International Migrant Household'. Baseline Effect does not include any controls and Controlled Effect has all the variables in Table A2 as controls. The identified set is bounded by $\beta$ of OLS model with controls and $\beta$ calculated given $R_{max} = 1.3 * \tilde{R}$ and $\delta = 1$. The last column shows the value of $\delta$ which would produce $\beta = 0$ given $R_{max} = 1.3 * \tilde{R}$. † indicates identified set does not include zero.* $*** \ p < 0.01$; $** \ p < 0.05$; $* \ p < 0.1$.

**Table A4**: First Stage Results for the IV Regressions

| | International Migrant Household |
|---|---|
| Taluk-religion-level migration network | 0.590*** (0.036) |
| Controls | Yes |
| Observations | 14325 |

*Note. Estimations from OLS regression is presented. The numbers in the parenthesis are the standard error clustered at the locality level. The dependent variable is whether the household has at least one international migrant. The main variable of interest is the IV, the proportion of households having an international migrant within the taluk from the same religion as that of the corresponding household. The control variables are those listed in Table A2 and district fixed effect.* $*** \ p < 0.01$; $** \ p < 0.05$; $* \ p < 0.1$

**Table A5**: Zero First Stage Sample

| | (1) Luxury House | (2) 5 least possessed | (3) VCI |
|---|---|---|---|
| Taluk-religion level - migration network | 0.172** (0.075) | 0.059 (0.052) | 0.053 (0.107) |
| Controls | Yes | Yes | Yes |
| Observations | 11815 | 11815 | 11815 |

*Note. Estimations from OLS regression are presented. The numbers in the parenthesis are the standard error clustered at the locality level. The dependent variables are as follows (i) column (1): possession of luxury house, (ii) column (2): possession of at least one out of five least possessed goods in the locality and (iii) column (3): visible consumption index (VCI). The main variable of interest is the IV, the proportion of households having an international migrant within the taluk from the same religion as that of the corresponding household. The control variables are those listed in Table A2.* $*** \ p < 0.01$; $** \ p < 0.05$; $* \ p < 0.1$

**Table A6**: Effect of transnational migration on visible consumption – IPWRA Estimates

| | Luxury House | 5 least possessed | VCI |
|---|---|---|---|
| | (1) | (2) | (3) |
| Migrant Household | 0.106*** | 0.089*** | 0.171*** |
| | (0.019) | (0.016) | (0.026) |
| Observations | 14325 | 14325 | 14325 |

*Note. Estimations from probit regression are presented in column (1) and (2) and for column (3), those from OLS regression are presented. The numbers in the parenthesis are the standard error clustered at the locality level. The dependent variables are as follows (i) column (1): possession of luxury house, (ii) column (2): possession of at least one out of five least possessed goods in the locality and (iii) column (3): visible consumption index (VCI). The main variable of interest is whether the household has at least one international migrant. All models have control variables that are listed in Table A2 and district fixed effect.* $*** \ p < 0.01$; $** \ p < 0.05$; $* \ p < 0.1$.

**Table A7**: Effect of remittances on visible consumption

| | Luxury House | | 5 least possessed | | VCI | |
|---|---|---|---|---|---|---|
| | (1) | (2) | (3) | (4) | (5) | (6) |
| | Probit | IV Probit | Probit | IV Probit | OLS | IV |
| Panel A | | | | | | |
| Remittances | 0.034*** | 0.098*** | 0.028*** | 0.055** | 0.018*** | 0.032** |
| | (0.003) | (0.026) | (0.003) | (0.025) | (0.002) | (0.015) |
| F-statistic | | 259.2 | | 259.2 | | 259.2 |
| Panel B | | | | | | |
| Plausibly Exogenous | | 0.049*** | | 0.020*** | | 0.054*** |
| | | (0.010) | | (0.007) | | (0.015) |
| Observations | 14325 | 14325 | 14325 | 14325 | 14325 | 14325 |

*Note. The numbers in the parenthesis are the standard error clustered at the locality level. The F-statistic refers to the Wald version of the Kleibergen and Paap rk-statistic. The dependent variables are as follows (i) column (1) and (2): possession of luxury house, (ii) column (3) and (4): possession of at least one out of five least possessed goods in the locality and (iii) column (5) and (6): visible consumption index (VCI). The main variable of interest is the hyperbolic sine of remittances received. The IV is the proportion of households having an international migrant within the taluk from the same religion as that of the corresponding household. Panel B gives the estimates assuming the IVs to be plausibly exogenous with uncertainty. All models have control variables that are listed in Table A2 and district fixed effect.* $*** \ p < 0.01$; $** \ p < 0.05$; $* \ p < 0.1$.

**Table A8**: Alternative Channels – Reference Group Change

| Dependent Variable: 5 least possessed | | | | | | |
|---|---|---|---|---|---|---|
| | (1) | (2) | (3) | (4) | (5) | (6) |
| Migrant Household | 0.661** | 0.730** | 0.735* | 0.656** | 0.742** | 0.787* |
| | (0.296) | (0.360) | (0.446) | (0.308) | (0.311) | (0.403) |
| Reference Group Income | | | | | | |
| Mean | | -0.058 | | | | 0.038 |
| | | (0.074) | | | | (0.088) |
| Median | | | -0.161 | | | -0.317 |
| | | | (0.319) | | | (0.411) |
| $90^{th}$ percentile | | | | 0.026 | | 0.136 |
| | | | | (0.131) | | (0.116) |
| Inequality | | | | | 1.170** | 0.705 |
| | | | | | (0.534) | (0.670) |
| F-statistic | 274.8 | 136.5 | 68.53 | 173.1 | 189.8 | 61.85 |
| Observations | 14325 | 14325 | 14325 | 14325 | 14325 | 14325 |

| Dependent Variable: VCI | | | | | | |
|---|---|---|---|---|---|---|
| | (1) | (2) | (3) | (4) | (5) | (6) |
| Migrant Household | 0.377** | 0.405* | 0.392 | 0.371** | 0.444** | 0.480** |
| | (0.178) | (0.214) | (0.262) | (0.183) | (0.190) | (0.243) |
| Reference Group Income | | | | | | |
| Mean | | -0.023 | | | | -0.008 |
| | | (0.040) | | | | (0.050) |
| Median | | | -0.030 | | | -0.154 |
| | | | (0.181) | | | (0.244) |
| $90^{th}$ percentile | | | | 0.053 | | 0.104 |
| | | | | (0.075) | | (0.080) |
| Inequality | | | | | 0.758** | 0.357 |
| | | | | | (0.312) | (0.359) |
| F-statistic | 274.8 | 136.5 | 68.5 | 173.1 | 189.8 | 61.9 |
| Observations | 14325 | 14325 | 14325 | 14325 | 14325 | 14325 |

*Note. Estimations from IV regressions are presented. The numbers in the parenthesis are the standard error clustered at the locality level. The F-statistic refers to the Wald version of the Kleibergen and Paap rk-statistic. The main variable of interest is whether the household has at least one international migrant. The dependent variables are as follows (i) Panel(A): possession of at least one out of five least possessed goods in the locality and Panel (B): visible consumption index (VCI). The IV is the proportion of households having an international migrant within the taluk from the same religion as that of the corresponding household. The control variables are those listed in Table A2.* $*** \ p < 0.01$; $** \ p < 0.05$; $* \ p < 0.1$.

**Table A9**: Mediation analysis exploring channels through which migration affect visible consumption

| Panel (A) - Mediator: Total Consumption Expenditure | | | |
|---|---|---|---|
| | (1) | (2) | (3) |
| | Luxury House | 5 least possessed | VCI |
| Mediation Effect | 0.010 | 0.013 | 0.038 |
| | [0.007, 0.013] | [0.010, 0.016] | [0.030, 0.046] |
| Direct Effect | 0.134 | 0.097 | 0.189 |
| | [0.113, 0.156] | [0.077, 0.118] | [0.151, 0.228] |
| Total Effect | 0.144 | 0.110 | 0.227 |
| | [0.122, 0.166] | [0.090, 0.130] | [0.188, 0.266] |
| Percentage Mediated | 6.8 | 11.7 | 16.7 |
| Observations | 14325 | 14325 | 14325 |
| **Panel (B) - Mediator: Median Income** | | | |
| | (1) | (2) | (3) |
| | Luxury House | 5 least possessed | VCI |
| Mediation Effect | 0.002 | 0.009 | 0.030 |
| | [-0.009, 0.012] | [-0.002, 0.018] | [0.010, 0.050] |
| Direct Effect | 0.142 | 0.102 | 0.197 |
| | [0.118, 0.167] | [0.079, 0.125] | [0.153, 0.241] |
| Total Effect | 0.144 | 0.110 | 0.227 |
| | [0.123, 0.166] | [0.090, 0.130] | [0.189, 0.266] |
| Percentage Mediated | 1.0 | 7.8 | 13.2 |
| Observations | 14325 | 14325 | 14325 |

*Note. Total effect captures the effect of migration on the visible consumption. Total effect is decomposed into mediation effect (explained by a specific mediator) and direct effect (not explained by the mediator). We consider mediator variables as indicated in the column heading. All models have control variables that are listed in Table A2 and district fixed effect. Robust standard errors clustered at the locality level are in parentheses. 95% confidence intervals are reported in the brackets.* $*** \ p < 0.01$; $** \ p < 0.05$; $* \ p < 0.1$.

**Table A10**: Effect of Income Unpredictability on Visible Consumption

| Panel A | | | |
|---|---|---|---|
| | (1)<br>Luxury House | (2)<br>5 least possessed | (3)<br>VCI |
| Household Income Unpredictability | 0.320*** | 0.442*** | 0.275*** |
| (1=above median SD of earnings from job) | (0.035) | (0.037) | (0.020) |
| Controls | Yes | Yes | Yes |
| Observations | 10480 | 10480 | 10480 |
| $\hat{\delta}\ for\ \beta = 0$ | 1.390 | 1.519 | 1.358 |

| Panel B | | | |
|---|---|---|---|
| | (1)<br>Luxury House | (2)<br>5 least possessed | (3)<br>VCI |
| Household Income Unpredictability | 0.320*** | 0.432*** | 0.308*** |
| (1=above mean SD of earnings from job) | (0.035) | (0.032) | (0.021) |
| Controls | Yes | Yes | Yes |
| Observations | 10480 | 10480 | 10480 |
| $\hat{\delta}\ for\ \beta = 0$ | 1.668 | 1.793 | 1.718 |

*Note. Estimations from probit regression are presented in columns (1) and (2) and for column (3), those from OLS regression are presented. The numbers in the parenthesis are the standard error clustered at the locality level. The dependent variables are as follows (i) column (1): possession of luxury house, (ii) column (2): possession of at least one out of five least possessed goods in the locality and (iii) column (3): visible consumption index (VCI). The main variable of interest, Household Income Unpredictability, in Panel A is a dummy variable that takes value one if earnings standard deviation of the income source of household exceeds the mean of the distribution of the same, and in Panel B, mean is used instead of median. The control variables are those listed in Table A2 and district fixed effects.* $*** \ p < 0.01$; $** \ p < 0.05$; $* \ p < 0.1$. *In the last row, the value of* $\hat{\delta}$, *from the coefficient stability (Oster) test, for which the coefficient estimate turns zero is given.*